\documentclass[twocolumn,showpacs,aps]{revtex4}
\usepackage{graphicx}
\begin{document}
\title{Effect of different site energies on polaronic properties}
\author{Jayita Chatterjee\footnote{ e-mail: moon@cmp.saha.ernet.in}
 and A. N. Das}

\affiliation{Theoretical Condensed Matter Physics Group \\
  Saha Institute of Nuclear Physics \\
1/AF Bidhannagar, Kolkata 700064, India}

\date{\today}

\begin{abstract}
Using the perturbation method based on a variational phonon basis
obtained by the modified Lang-Firsov (MLF) transformation,
the two-site single polaron Holstein model is studied in presence of
a difference in bare site energies ($\epsilon_d$=$\epsilon_2$-$\epsilon_1$).
The polaronic ground-state wave function is calculated up to the fifth 
order of perturbation. The effect of $\epsilon_d$ (acts as a site-energy
disorder) on the polaron 
crossover, polaronic kinetic energy, oscillator wavefuncion and
polaron localization are studied. 
Considering a double-exchange Holstein model with finite $\epsilon_d$, 
role of disorder on the properties of the double-exchange system
is also discussed.
\end{abstract}

\pacs{~71.38. +i, 63.20.kr}  
\maketitle

\begin{center}
{\bf I. Introduction}
\end{center}
\vskip 0.3cm

Study of narrow band electronic systems with strong electron-phonon ($e$-ph) 
interaction has long been an active research area in condensed 
matter physics. The field has drawn renewed interest following
evidence of polaronic charge carriers in underdoped high-$T_c$ 
cuprates \cite{hightc}, manganites \cite{zhao},
and organic superconductors.
The one-dimensional polaron problem is also relevant in semiconductor 
physics, quantum dots \cite{lep} and linear conjugated organic polymer
conductors \cite{ssh2}. 
The simplest model for studying polarons  
is the Holstein model \cite{Hol} where an electron in a narrow band 
interacts locally with optical phonons. For large $e$-ph coupling the 
polaron is a small polaron with high effective mass, while
for small coupling it becomes a large polaron having a
much lower effective mass for a finite adiabatic parameter. 
The crossover from a large to a small polaron and the corresponding
change in the polaronic properties in the ground state have been 
studied for the Holstein model by different groups \cite {Jeck,Romero,
Trugman,Barisic} using various methods enlightening 
our understanding in this field. However studies on the nature and
properties of polarons in presence of disorder are few and need 
much more attention.
The imperfections or disorder may play an important role in 
complex materials (high-$T_c$ oxides, manganites etc.) where 
signatures for polaronic carriers are found. 
Recently the small polaron concept has been used to explain
the charge motion in DNA where the electronic band is very narrow 
and the presence of different kind of molecular units induces large
disorder potential \cite{DNA}. 
 
In absence of any disorder, translational symmetry ensures that 
the polaronic ground state is delocalized however large is the 
$e$-ph coupling strength provided other
parameters (electronic hopping, phonon frequency) are finite. 
The large to small polaron crossover is a continuous one 
\cite{Trugman}, which is consistent with the ground state properties being 
analytic functions of $e$-ph coupling\cite{L}.   
Localization requires a breakdown of the translational invariance 
which may be achieved through randomness of the site potential 
or hopping. The effect of site diagonal disorder on polaronic 
properties has been addressed by some authors.
Shinozuka and Toyozawa \cite{ST} studied disorder induced self-trapping
in a tight binding model in which the local site energies are randomly 
distributed between two values and found that the 
exciton-lattice interaction acts with the disorder to produce severe 
localization associated with a self-trapped exciton. In his study
the lattice vibration was treated as classical oscillators.
Bronold $et.~al$ \cite{BF2} studied similar model but with an 
infinite coordination number within the dynamical coherent potential 
approximation. However the limitation of the coherent potential 
approximation is that it cannot fully distinguish between localized and 
itinerant states. Bronold and Fehske \cite{BF} improved the method
to overcome the above shortcoming. They followed statistical dynamical mean 
field theory to predict localization of small polarons by extremely
small disorder. However a proper study of the effect of disorder
on the polaron crossover is not made in any of the above investigations. 
In the present work we consider a two-site cluster with different
site energies. This is the minimal system to study the competition
of the inter-site electronic hopping with the localization  
induced by the combined effect of the $e$-ph coupling and the 
site energy disorder. Difference in site energies would 
remove the two-fold degeneracy of the system in the absence of hopping
and would tend to localize the electron in the lower potential site.
{\it For convenience we will refer to the difference in site energies 
as disorder strength because it partly mimics 
the role of disorder in larger systems}. Another aspect of choosing
a two-site Holstein model is that almost exact results may be 
obtained for such a system with the perturbation method \cite{JC2000}
using a modified Lang-Firsov (MLF) basis \cite{DS}. For the Holstein
model the interaction is very short-ranged and the essential physics
relating polaronic behavior for a larger system is similar to that in a 
two-site system. In studying a Hubbard-Holstein model similar conclusion 
has been reached in Ref. \cite {Takada}.
The relevance of studying a two-site system in the context of Holstein 
and Holstein-double exchange models has been discussed in details 
in Ref. \cite {Capon}. 

 In Sec. II we discuss the formalism and perturbation calculations.
In Sec. III we present the results obtained by
MLF method and discuss the role of the
disorder strength on the polaron crossover and the kinetic 
energy of the system. The localization of the polaron and ground
state polaronic wavefunction will also be discussed. Extension of this model
to the double-exchange model is included in Sec. IV.
Finally, a summary of the results are presented in Sec. V.   

\vskip 1.0cm

\begin{center}
{\bf II. Formalism }
\end{center}
\vskip 0.3cm

The two-site single-polaron Hamiltonian is
\begin{eqnarray}
H &=&  \epsilon_1 n_1 +\epsilon_2 n_2 - \sum_{\sigma}
t (c_{1 \sigma}^{\dag} c_{2 \sigma} + c_{2 \sigma}^{\dag} c_{1 \sigma}) 
\nonumber \\
&+& g \omega  \sum_{i,\sigma}  n_{i \sigma} (b_i + b_i^{\dag})
+  \omega \sum_{i}  b_i^{\dag} b_i
\end{eqnarray}
where $i$ =1 or 2, denotes the site. $\epsilon_{1}$ and $\epsilon_{2}$ 
are the bare site-energies at site 1 and 2, respectively.
$c_{i\sigma}$ ($c_{i\sigma}^{\dag}$)
is the annihilation (creation) operator for the electron with spin
$\sigma$ at site $i$ and $n_{i \sigma}$ (=$c_{i\sigma}^{\dag} c_{i\sigma}$)
is the corresponding number operator, $g$ denotes the on-site $e$-ph coupling
strength, $t$ is the usual hopping integral.
$b_i$ and $b_{i}^{\dag}$ are the annihilation and
creation operators, respectively, for the phonons corresponding to
interatomic vibrations at site $i$ and $\omega$ is the phonon frequency.
This Hamiltonian has spin degeneracy for the one electron case so
the spin index is redundant.

 Introducing new phonon operators $a=~(b_1+b_2)/ \sqrt 2$ and
$d=~(b_1-b_2)/\sqrt 2 $, the Hamiltonian is separated into two parts
($H=H_d + H_a)$ :
\begin{eqnarray}
H_d &=&  \epsilon_1 n_{1} +\epsilon_2 n_2
- t (c_{1}^{\dag} c_{2} + c_{2}^{\dag} c_{1})
\nonumber \\
&+& \omega  g_{+} (n_1-n_2) (d + d^{\dag}) 
+  \omega  d^{\dag} d \\
{\rm and}~~ H_a &=&  \omega \tilde{a}^{\dag}\tilde{a} - \omega n^2 g_{+}^2
\end{eqnarray}
where $g_{+}=g/\sqrt 2$, $\tilde{a}=a +ng_{+}$.
The $a$-oscillator couples only with the total number of electrons 
$n(=n_1+n_2)$, which is a constant of motion. Consequently $H_a$ 
describes just a shifted oscillator, while $H_d$ represents an effective 
$e$-ph system where phonons couple with the electronic degrees of freedom.
For a perturbation method it is desirable to use a basis
where the major part of the Hamiltonian becomes diagonal.
When the hopping ($t$) is appreciable compared to the polaron 
energy ($g^2 \omega$), a retardation between the electron and 
associated lattice distortion sets in. The retardation is important
for low and intermediate values of $g$; it produces a spread in 
the polaron size, the resultant polaron is a large one. 
In Ref. \cite {JC2000} we have shown that the MLF perturbation method 
works much better than the Lang-Firsov (LF) method
for a large region of parameter space where the retardation is important.
In the strong-coupling limit the MLF method reduces to the LF method 
and it works well there. 
We use the MLF transformation where the lattice deformations
produced by the electron are treated as variational parameters
\cite{DS,DC}. For the present system, $\tilde{H_d} = e^R H_d e^{-R}$
where $R =\lambda (n_1-n_2) ( d^{\dag}-d)$ and $\lambda$ is a
variational parameter describing the displacement of the $d$
oscillator.

 The transformed Hamiltonian is then obtained as
\begin{eqnarray}
\tilde{H_d}&=&\omega  d^{\dag} d + 
(\epsilon_1- \epsilon_p) n_{1} + (\epsilon_2- \epsilon_p) n_{2} 
\nonumber \\
&-&t ~[c_{1}^{\dag} c_{2}~ \rm{exp}(2 \lambda (d^{\dag}-d))   
\nonumber \\
&+& c_{2}^{\dag} c_{1} \exp(-2 \lambda (d^{\dag}-d))]
\nonumber\\
&+& \omega  (g_{+} -\lambda) (n_1-n_2) (d + d^{\dag}) \\
{\rm where} ~\epsilon_p &=&  \omega ( 2 g_{+} - \lambda) \lambda. 
\nonumber
\end{eqnarray}

For the single polaron problem we choose the basis set,
\begin{eqnarray}
|+,N \rangle &=&  (a_1 c_{1}^{\dag} 
+a_2  c_{2}^{\dag}) |0\rangle_e  |N\rangle_{{\rm ph}} \nonumber \\
|-,N \rangle &=&  (a_2 c_{1}^{\dag} -a_1  c_{2}^{\dag}) |0\rangle_e  
|N\rangle_{{\rm ph}}
\end{eqnarray}
where $|+\rangle$ and $|-\rangle$ denote the
electronic states and $|N\rangle$ denotes the $N$th excited oscillator
state in the MLF phonon basis. The normalization condition requires
$a_1^2 + a_2^2 =1$.
The unperturbed part of the Hamiltonian is chosen as
\begin{eqnarray}
H_0&=&\omega  d^{\dag} d + 
(\epsilon_1- \epsilon_p) n_{1} + (\epsilon_2- \epsilon_p) n_{2} 
\nonumber \\
&-&t_e ~(c_{1}^{\dag} c_{2}+c_{2}^{\dag} c_{1})
\end{eqnarray}
where $t_{e}=t~\rm{ exp}{(-2\lambda^2)} $.
The remaining part $H_1 = (\tilde{H_d}-H_0)$ is treated as a perturbation.
The states $|\pm,N\rangle$ are the eigenstates of the unperturbed 
Hamiltonian ($H_0$) for $r=a_1 /a_2= [ \epsilon_d
+  \sqrt{{\epsilon_d}^2 +4 {t_e}^2 }]/2t_e$ where 
$\epsilon_d = \epsilon_2 - \epsilon_1$. The corresponding
eigen energies are given by
\begin{equation}
 E_{\pm,N}^{(0)}= N\omega 
 + \frac{(\epsilon_1 +\epsilon_2)}{2} -\epsilon_p 
\mp \frac{1}{2} \sqrt{(\epsilon_1-\epsilon_2)^2 +4 {t_e}^2 }
\end{equation}
The state $|+,0\rangle$ has the
lowest unperturbed energy, $ E_0^{(0)}=\epsilon_1 +\frac{\epsilon_d}{2}
-\epsilon_p-\frac{1}{2} \sqrt{{\epsilon_d}^2 +4 {t_e}^2 }$.

 The general off-diagonal
matrix elements between the states $|\pm,N \rangle$ and $|\pm,M \rangle$
are calculated for $(N-M)>0$ as follow :\\
for even $(N-M)$, 
\begin{eqnarray}
\langle N,\pm|H_1|\pm, M \rangle &=& \mp t_e \frac{2r}{(1+r^2)}\\
\langle N,\pm|H_1|\mp,M \rangle &=& -t_e \frac{(1-r^2)}{(1+r^2)}
\end{eqnarray}
for odd $(N-M)$,
\begin{eqnarray}
\langle N,\pm|H_1|\pm, M \rangle &=& 
\mp \sqrt{N}\omega (g_+ -\lambda) 
 \frac{(1-r^2)}{(1+r^2)} \delta_{N,M+1} \\
\langle N,\pm|H_1|\mp,M \rangle &=& \pm t_e  
+ \sqrt{N}\omega (g_+ -\lambda)\frac{2r}{(1+r^2)}\delta_{N,M+1} \nonumber\\
\end{eqnarray}
It may be noted that for the ordered case ($\epsilon_d$=0, hence $r$=1)
the off-diagonal matrix elements 
$\langle N,\pm|H_1|\pm, M \rangle$ are nonzero only for even $(N-M)$
while $\langle N,\pm|H_1|\mp, M \rangle$ are nonzero only for odd $(N-M)$.

 Now one has to make a choice of $\lambda$ so that the perturbation
corrections are small and perturbative expansion becomes convergent.
Here we will follow the procedure of our previous work \cite{JC2000}
to find out the variational phonon basis as a function of $e$-ph coupling. 
Minimizing the unperturbed ground state energy $E_0^{(0)}$
with respect to $\lambda$ we obtain
\begin{equation}
\lambda=\frac{ \omega g_+}{\omega + \frac{4{t_{e}}^2}{\sqrt{{\epsilon_d}^2 + 
4 {t_e}^2 }  }  }
\end{equation}

The perturbation corrections of different orders to the ground-state
energy and the wavefunction may be calculated using the general
off-diagonal matrix elements in Eqs.(8-11).
The ground-state wave function in the MLF basis may be written as,
\begin{eqnarray}
|G \rangle
 &=& \frac{1}{\sqrt{N_G}} \left[ |+,0\rangle
 +\sum_{N=1,2,..}c_{N}^{+} |+,N \rangle \right. \nonumber \\
 &+& \left. \sum_{N=1,2,..}c_{N}^{-} |-,N \rangle \right]
\end{eqnarray} 
The coefficients $c_{N}^{\pm}$ are obtained in terms of the
off-diagonal matrix elements and unperturbed energies  
and $N_G$ is the normalization factor.
  
 The correlation functions involving the charge and the lattice deformations
$\langle n_1 u_{1}\rangle_{0}$ and $\langle n_1 u_{2}\rangle_{0}$,
where $u_1$ and $u_2$ are the lattice deformations at sites 1 and 2
respectively, produced by an electron at site 1, are the standard
measure of polaronic character and indicate the strength of polaron
induced lattice deformations and their spread. Following Ref.\cite{JC2000},
the correlation functions may be written as
\begin{eqnarray}
\langle n_{1} u_{1} \rangle&=&\frac{1}{2}
\left[-(g_{+} + \lambda) \frac{B_0}{N_G} + \frac{A_0}{N_G}\right] \\
\langle n_{1} u_{2} \rangle&=&\frac{1}{2}
\left[-(g_{+} - \lambda) \frac{B_0}{N_G} -  \frac{A_0}{N_G}\right] \\
{\rm where}~~A_0 &\equiv& \langle G |n_1(d+d^{\dag})|G \rangle =
\frac{2}{(1+r^2)} (  r c_1^{-} +  r^{2} c_{1}^{+}) \nonumber\\
&+& \sum_{N=1}^{\infty}  \sqrt{N+1} \frac{2}{(1+r^2)}
\left[  r^{2} c_N^{+} c_{N+1}^{+} \right.  \nonumber\\
&+& \left. r (c_{N}^{-}c_{N+1}^{+} + c_{N}^{+}c_{N+1}^{-} )
+   c_{N}^{-}c_{N+1}^{-} \right] \nonumber\\
{\rm and}~~B_0 &\equiv& \langle G |n_1|G \rangle 
= \frac{r^2}{(1+r^2)}+ \sum_{N=1}^{\infty} \frac{1}{(1+r^2)} \nonumber\\
&& \left[ 2r c_N^{-} c_{N}^{+} 
+  r^2 |c_N^{+}|^2 +|c_N^{-}|^2 \right] \nonumber
\end{eqnarray}
The static correlation functions in Eqs. (14) and (15)
are calculated for different $\epsilon_d$ to examine
the role of the disorder on the polaron crossover.

  The kinetic energy of the system in the ground state is 
obtained as 
\begin{eqnarray}
E_{K.E}&=& <G|H_t |G> \nonumber\\
&=& \frac{1}{N_G} \left[-~ 2t_e a_1 a_2+ 2 \sum_{e=\pm} \sum_{N \neq 0} 
c_{N}^{e}\langle e,N|H_t|0,+\rangle \right.
\nonumber\\
& & \left.+ \sum_{e,e^{\prime}=\pm}~~ \sum_{N,M \neq 0}
c_{N}^{e}c_{M}^{e^{\prime}} 
\langle e,N|H_t|M,e^{\prime} \rangle \right]
\end{eqnarray}
where $H_t=-t [c_{1 }^{\dag} c_{2 }~ \exp(2 \lambda (d^{\dag}-d))
+ c_{2}^{\dag} c_{1 }\exp(-2 \lambda$\\
$ (d^{\dag}-d))]$ is the 
kinetic energy operator in the MLF basis.
 The occupation number $n(k)$ of the charge carrier for the ground state
within MLF method is also calculated as
\begin{eqnarray}
n_{0,\pi} &=& \frac{1}{2} (c_{1 }^{\dag}\pm c_{2 }^{\dag})
(c_{1}\pm c_{2}) \\
 \langle G |n_{0,\pi}|G \rangle &=&
\frac{1}{2}\left[ 1  \pm \frac{E_{K.E} }{t}\right] 
\end{eqnarray}
where $0,\pi$ denotes the values of the wave vector $k$.
Therefore, it basically reflects the nature of the kinetic energy.

The ground-state wave function for the d-oscillator is obtained 
from Eq. (13) by using the wavefunction for the N-th excited 
(MLF-displaced) harmonic oscillator for $|\pm, N\rangle$. 
\begin{equation}
\psi_{n}(x) = \langle x|n \rangle =\frac{1}{\pi^{1/4} \sqrt{ 
2^{n} n!} }e^{-(x-x_0)^2/2} H_{n} (x-x_0)
\end{equation}
where $H_{n}(x)$ is the Hermite polynomial of degree n and
$x_0$ is the displacement of the oscillator due to the MLF
transformation.

 To study the localization effect due to disorder in
site energy we calculate $|_e\langle 1|G\rangle|^2$ and
$|_e\langle 2|G\rangle|^2$. These are the probabilities that
the polaron in the ground state ($|G\rangle$) lies
at site 1 and site 2, respectively.
\begin{eqnarray}
|_e\langle 1|G\rangle|^2 &=&
\frac{1}{N_G} \left[ |a_1|^{2} + \sum_{N \neq 0} |
a_1 c_N^{+} +a_2 c_N^{-} |^2 \right]  \\
|_e\langle 2|G\rangle|^2 &=&
\frac{1}{N_G} \left[ |a_2|^{2} + \sum_{N \neq 0} |
a_2 c_N^{+} -a_1 c_N^{-} |^2 \right] 
\end{eqnarray}

\begin{center}
{\bf III. Results and discussions}
\end{center}
\vskip 0.3cm
	
\begin{figure}
\centering
\includegraphics[scale = 0.35,angle=270]{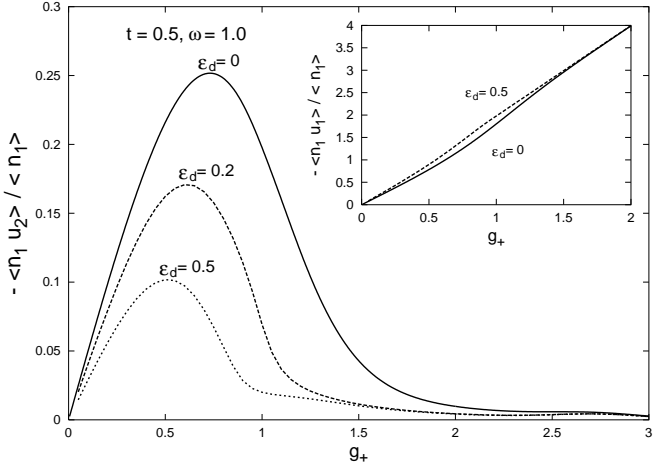}
\caption{Plot of the correlation function $\langle n_1u_2 \rangle$
versus $g_{+}$ for $t/\omega =0.5$
for different values of $\epsilon_d$. Inset: 
on-site correlation function $\langle n_1u_1 \rangle$.
}
\end{figure}

In this paper all the results are derived by calculating 
the ground-state wavefunction up to the fifth order of perturbation. 
For the ordered case a comparison of our MLF-perturbation results 
\cite {JC2000}
with the exact results by Rongsheng $et.~ al$ \cite{Rongsheng} shows 
that the MLF method up to the fifth order gives exact 
results for $t=0.5$ (in a scale where $\omega$=1) whereas 
for higher values of $t$ (=1.1 and 2.1) very accurate results are produced 
by the MLF perturbation method for both strong and weak coupling regions.
However in a narrow region of intermediate coupling the perturbation results
for high values of $t$ deviate from the exact results. 
In this paper we present the results mainly for the nonadiabatic
regime ($t/\omega \le 1.0$) for
which the convergence of the MLF-perturbation
series for both ordered and disordered cases is found to be very good
in the entire region of the $e$-ph coupling strength.
For the ordered case, the convergence in energy and correlation functions
has already been reported in Ref.\cite{JC2000} and that for the wave function
in Ref.\cite{IJ01}.

 In Fig. 1 we plot the variation of the correlation functions
$\langle n_1u_1 \rangle$ and $\langle n_1u_2 \rangle$
with $g_{+}$ for $t/\omega =0.5$ and for different values of 
the disorder strength $\epsilon_d$.
For intermediate coupling $\langle n_1u_2 \rangle$ has appreciable
values and $\langle n_1u_1 \rangle$/$\langle n_1 \rangle$ (shown in inset)
deviates downwards from its small polaronic LF value of 2$g_+$.
These are the signatures of retardation and directly show
the spread up of the polaron.
With increasing $\epsilon_d$ the value of $\langle n_1u_2 \rangle$ 
decreases and $\langle n_1u_1 \rangle$ shows less deviation from 
its LF value. This points to a reduced retardation effect with increasing 
disorder strength.         

\begin{figure}[t]
\centering
\includegraphics[scale = 0.5,angle=0]{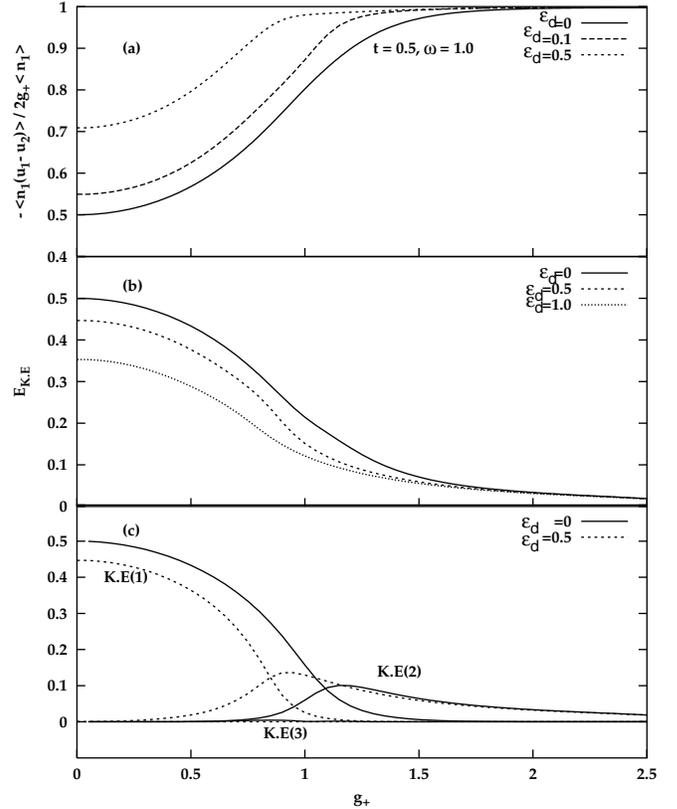}
\caption{(a) The variations of $\chi=\langle n_{1} (u_{1}-u_{2}) \rangle/2g_+\langle
n_1\rangle $ with $g_+$ for
different values of $\epsilon_d$ for $t/\omega=0.5$.
(b) The variation of the kinetic energy $E_{K.E}$
with $g_+$ for $t/\omega$=0.5 and $\epsilon_d$=0, 0.5
and 1.0. 
(c) Different parts of the kinetic energy (see the three terms in Eq (16)
referred as K.E(1), K.E(2) and K.E(3), respectively)
vs. $g_+$ for different $\epsilon_d$.
}
\end{figure}
 The polaron crossover from a large to a small polaron may be studied 
by the correlation function 
$\chi=\langle n_{1} (u_{1}-u_{2}) \rangle/2g_+\langle n_1\rangle$. 
In the small polaron limit  the retardation effect is negligible and  $\chi$ 
gets its standard LF value (=1). For a larger size polaron the value 
of $\chi$ is lower. 
In Fig. 2(a) we plot $\chi$ as a function of $e$-ph coupling strength.
It is seen that the size of the polaron becomes more localized 
with increasing disorder strength. The polaron crossover is 
continuous, but a change in the curvature of $\chi$ vs. $g_+$ plot
is observed at a point in the crossover region. This point (in 
$g_+$ space) where
the curvature in $\chi$ vs. $g_+$ plot changes may be taken as 
the crossover point. The crossover point shifts to lower value 
of $g_+$ as $\epsilon_d$ increases. Thus the disorder favors 
formation of small polarons.    
In Fig. 2(b) we plot the kinetic energy of the polaron as a function
of $g_+$. The kinetic energy is suppressed in presence of
disorder in the range from low to intermediate values of $g_+$. 
But disorder has no significant effect on the kinetic energy for 
strong coupling where the polarons are small. The kinetic energy in 
Eq. (16) has three terms. The first
term is the contribution to the kinetic energy from the unperturbed
ground state which is significant for low to intermediate couplings.
The second term is the contribution from the matrix
elements of $H_t$ connecting the unperturbed ground state to higher 
phonon states. The third term is due to the hybridization of higher
phonon states through $H_t$. We plot these contributions separately
as a function of $g_+$ in Fig. 2(c) for $\epsilon_d$=0 and 0.5.
It is found that the first contribution, which is dominant
in the range from low to intermediate couplings and proportional
to the product of the coefficients $a_1$ and $a_2$ (see Eq.(16)), is suppressed 
substantially with increasing disorder strength. 
For strong-coupling regime the contribution to the kinetic energy 
comes only from the second term of Eq.(16) and that one is not affected
by disorder in this region. It may be mentioned that contribution
from the third 
term is too small to show up in the figure for the range of our study.    

\begin{figure}[]
\centering
\includegraphics[scale = 0.52,angle=0]{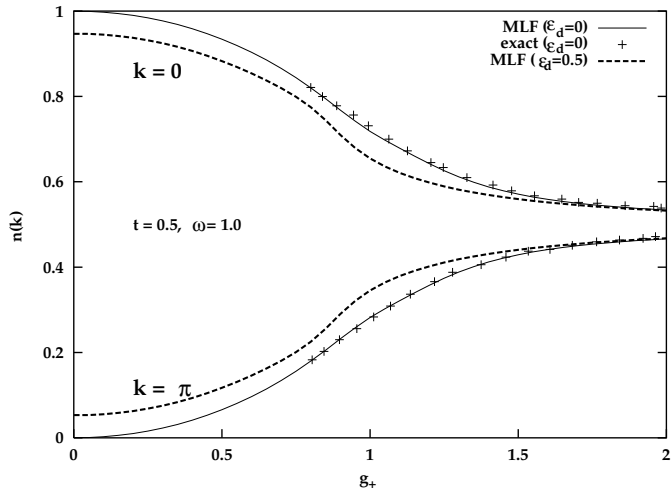}
\caption{The wave vector dependence of the occupation number
$n(k=0,\pi)$ of the charge carrier for the ground state with $e$-ph coupling
within MLF
method for $\epsilon_d$=0 and 0.5. Exact results (for $\epsilon_d$=0)
from Ref. \cite{MR} is also plotted for comparison. 
}
\end{figure}
In Fig. 3, we have plotted the occupation number $n(k)$ of the polaron 
for the ground state with $e$-ph coupling.
The difference between $n(k$=0) and $n(k$=$\pi)$ depends directly 
on the delocalization of the polaron (see Eq (18)). In absence of
hopping this difference vanishes. With increasing $e$-ph coupling
the difference between $n(k)$ reduces as the kinetic energy is more 
and more suppressed. For the disordered case and in the range from
weak to intermediate coupling, the difference in occupations reduces
further owing to disorder-induced suppression of the kinetic energy.
In the strong-coupling regime, $n(k)$
does not show any change with $\epsilon_d$ because 
the kinetic energy is almost independent of $\epsilon_d$ in this region.
The exact results for the occupation number \cite{MR} for the ordered 
case ($\epsilon_d$=0) are also shown in Fig. 3 for comparison with our 
MLF results.

\begin{figure}[]
\centering
\includegraphics[scale = 0.5,angle=0]{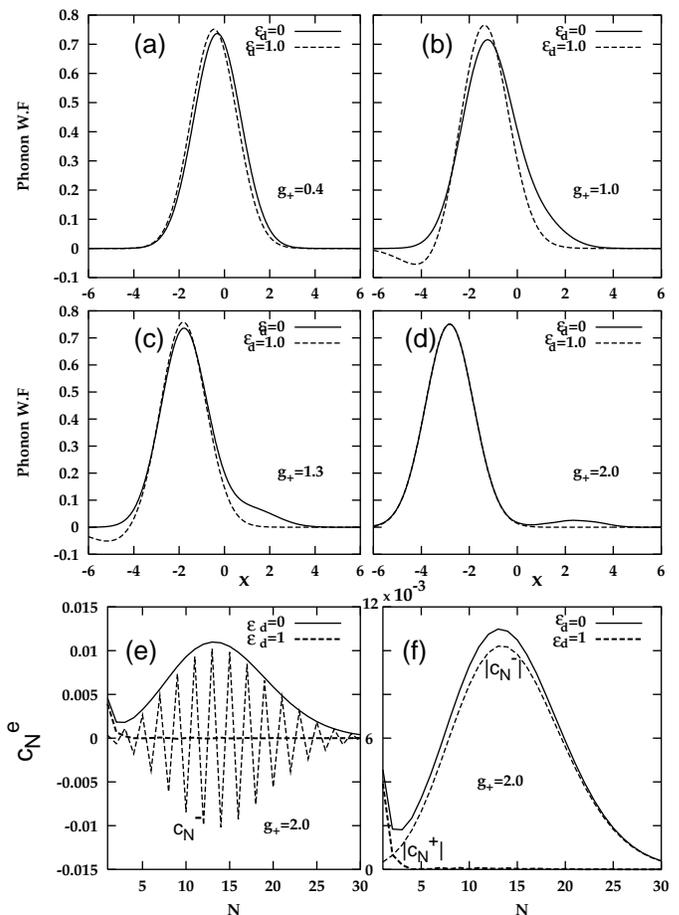}
\caption{Oscillator wave function (not normalized) vs. $x$ 
with $t/\omega=0.5$ for $\epsilon_d=0$ and 1.0
for (a) $g_+=0.4$, (b) $g_+=1.0$, 
(c) $g_+=1.3$ and (d) $g_+=2.0$. 
(e) The coefficients $c_{N}^{e}$ (see Eq. (13)) of the ground state
wave function  with phonon number $N$ for $t/\omega$= 0.5 and $g_+$=2
for $\epsilon_d$=0 and 1.0.
Solid curve gives the value of $c_{N}^{-}$ for odd $N$ and
$c_{N}^{+}$ for even $N$ for the ordered case.
For disordered case the thick dashed curve shows $c_{N}^{+}$
and the thin dashed curve shows $c_{N}^{-}$.
(f) Magnitude of the coefficients $(|c_N^{e}|)$ are 
shown for the same case as (e).
}
\end{figure}

\begin{figure}[]
\centering
\includegraphics[scale = 0.5,angle=0]{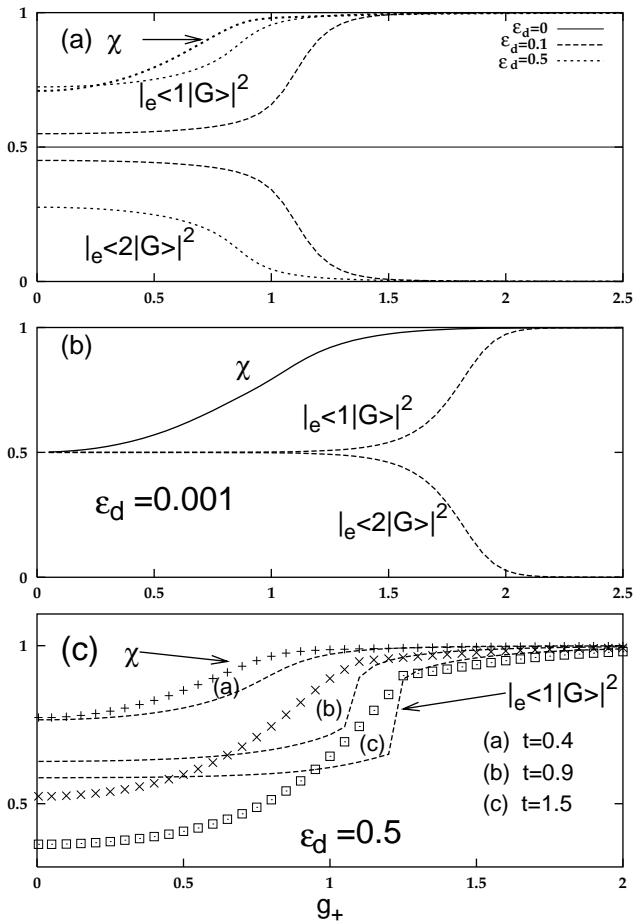}
\caption{Plots of $|_e\langle 1 |G\rangle|^2$
and $|_e\langle 2 |G\rangle|^2$
versus $g_+$ for $t/\omega =0.5$. $|_e\langle 1 |G\rangle|^2$
and $|_e\langle 2 |G\rangle|^2$ are the probabilities of site 1
and site 2 being occupied by the electron, respectively.
(a) The solid line is for the ordered case ($\epsilon_d$=0)
where both occupancy are same, thick dashed line for $\epsilon_d$=0.1 and
thin dashed line for $\epsilon_d$=0.5.
(b) The same curves for $\epsilon_d$=0.001. Polaron crossover
($\chi$) is also shown. (c) Plots of $\chi$ and $|_e\langle 1 |G\rangle|^2$
for different values of $t$. $\chi$ is denoted by different
symbols and $|_e\langle 1 |G\rangle|^2$ by broken line.
}
\end{figure}

 In Figs. 4(a-d) we have shown the ground-state wave functions 
of the $d$ oscillator for different values of $e$-ph coupling
for both the ordered and disordered cases 
considering that the electron is located at site 1.
The results for the ordered case have been discussed in
previous works \cite{MR, JC2000, Rongsheng}. The main changes, 
which are observed for the disordered case ($\epsilon_d$=1.0) in
the antiadiabatic regime, are as follow:\\
(i) for weak coupling ($g_+=0.4$) the wave function, which shows 
displaced Gaussian-like single peak, is slightly more shifted; 
(ii) for intermediate coupling ($g_+=1.3$), 
the additional shoulder, which appears (at the right side of the main 
peak) for the ordered case, is absent in presence of disorder; 
(iii) for strong coupling ($g_{+}=2.0$) though the wavefunction around the 
main peak is identical for both cases, 
the additional small broad peak, observed for the ordered case, 
is not found for the disordered case. The first feature indicates that
the on-site polaronic deformation is larger for the disordered case;
while the second feature may be due to the reduced retardation effect.
The third feature, i.e., disappearance
of the additional broad peak for the disordered case, has
a contradiction with the behavior of the kinetic energy.
Both this broad additional peak and the kinetic energy in the strong-coupling
regime are determined by the coefficients $c_N^{\pm}$ in Eq.(13).
A logical question then arises why the broad peak in the oscillator
wave function is modified by disorder while the kinetic energy in that region
is unaffected. In Fig. 4(e)  
we plot $c_N^{\pm}$ versus $N$ for $g_+=2.0$ for both ordered and
disordered cases. For the ordered case
$c_N^{-}$ is nonzero for odd $N$, while $c_N^{+}$ is nonzero for
even $N$. For the disordered case both the coefficients
$c_N^{-}$ and $c_N^{+}$
are nonzero for any $N$. For strong coupling, the reduced polaronic
hopping makes the value of $r$ (see the expression of $r$ after Eq.(6))
very large in case of disorder. As a consequence
we find that $c_N^{+}$ takes negligible
value except for $N=1$, while $c_N^{-}$ is appreciable for both odd 
and even $N$. $c_N^{-}$ changes sign alternately with $N$. 
However, the magnitude of the coefficients ($|c_N^{e}|$)
for any $N$ do not change much with disorder strength in the
strong-coupling regime (shown in Fig. 4(f)).
In this region we observe that the 
kinetic energy essentially depends on the magnitude of $c_N^e$
rather than on its sign,
hence it remains almost unaffected by disorder. However, the oscillator
wavefunction which is given by
\( \sum_{N,e}c_{N}^{e} |N \rangle_{{\rm ph}} \),
depends also on the sign of $c_N^e$ and is affected by disorder.       
 
 In Fig. 5 we plot $|_e\langle 1|G\rangle|^2$ and
$|_e\langle 2|G\rangle|^2$, which give the probabilities that
the polaronic charge carrier in the ground state ($|G\rangle$) 
lies at site 1 and site 2, respectively. For the ordered case
these probabilities are same (=0.5) for any value of the coupling. 
With increasing disorder strength these probabilities differ
from 0.5; the site with lower site-energy has higher probability
of being occupied than the other. This difference in the occupancy
increases rapidly during polaron crossover as a consequence of
rapid suppression of the polaronic hopping ($t_e$) 
with increasing $g_+$ in the crossover region.
For strong-coupling regime $|_e\langle 1|G\rangle|^2$ becomes
almost 1 while $|_e\langle 2|G\rangle|^2$ approaches to zero showing
localization of the polaron at site 1.
This localization as well as the correlation function $\chi$ 
are shown in Fig. 5(a) and 5(b) for $\epsilon_d$ =
0.5, 0.1 and 0.001 to show that the polaronic
crossover precedes the localization for $t/\omega=0.5$.
In Fig. 5(c) similar plots are presented for different values of
$t/\omega \le 1.5$ and $\epsilon_d=0.5$. With increasing $t$
the polaron crossover and the localization occur at higher values of 
$e$-ph coupling as expected. However, the localization is abrupt
but continuous for larger values of $t$ while the polaron crossover
is still very smooth for $t/\omega \le$1.5.


 It may be mentioned that when $t/\omega \ge$2 and
$\epsilon_d \ge$0.5 we find that the polaron crossover and the 
localization occur almost simultaneously. However, for such large values 
of $t$ our perturbation method does not converge properly
in a small range around intermediate coupling where polaron crossover
takes place.

 In Fig. 6 we plot $\chi$ and $|_e\langle 1 |G\rangle|^2$ as 
a function of disorder strength for different values of $e$-ph
coupling with $t/\omega$=0.5. For very weak coupling ($g_+=0.1$) 
the polaron crossover is induced by disorder potential 
and is very smooth. Here the localization and the polaron crossover 
occur almost simultaneously but none of them is complete even 
at $\epsilon_d=2$.
For $g_+=0.4$, which is also in the weak coupling range, the localization
follows the polaron crossover and a large value of disorder strength 
is required for localization. 
For intermediate coupling ($g_+=1$) the localization takes place in 
the small polaron region and 95$\%$ of the localization is achieved within
$\epsilon_d /t \sim 1$.  For higher values of coupling the polaron would 
be a small polaron even in absence of disorder and localization would 
occur for a small value of disorder strength. 
Our study points out that when the $e$-ph coupling is in 
the intermediate range, localization may be achieved with a disorder 
strength of the order of half (electronic) bandwidth of the system; this
has also been pointed out in Ref.\cite{Edwards} in the context of manganites.
For small $e$-ph coupling, disorder strength larger than
the bandwidth is required for localization.

\begin{figure}[]
\centering
\includegraphics[scale = 0.35,angle=270]{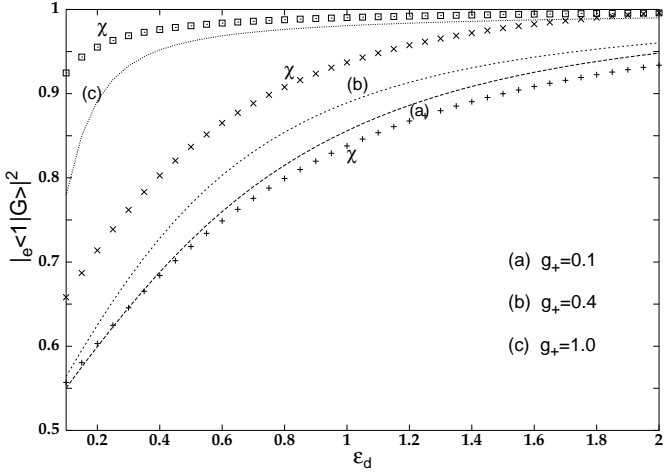}
\caption{Plots of $\chi$ and $|_e\langle 1 |G\rangle|^2$
versus $\epsilon_d$ for different values of $g_+$ with $t/\omega =0.5$. 
$\chi$ is denoted by different
symbols and $|_e\langle 1 |G\rangle|^2$ by broken line.
}
\end{figure}

\vskip 0.5cm
\begin{center}
{\bf IV. Two-site Holstein model with double exchange: effect of disorder}
\end{center}
\vskip 0.3cm

The relevant Hamiltonian for studying a two-site
double exchange Holstein model in presence of antiferromagnetic
interaction between core spins is given by \cite{EPJB,Capon}
\begin{eqnarray}
H &=&  \epsilon_1 n_1 +\epsilon_2 n_2 - \sum_{\sigma}
t~ cos(\theta/2) (c_{1 \sigma}^{\dag} c_{2 \sigma}
+ c_{2 \sigma}^{\dag} c_{1 \sigma}) \nonumber \\
&+& g \omega  \sum_{i,\sigma}  n_{i \sigma} (b_i + b_i^{\dag})
+  \omega \sum_{i}  b_i^{\dag} b_i + J S_1.S_2
\end{eqnarray}
where $S_1$, $S_2$ represent the local core spins (for manganites it is
the spin of $t_{2g}$ electrons) at sites 1 and 2, respectively and
$\theta$ is the angle between them.
$J$ is the superexchange antiferromagnetic interaction
between the neighbouring core-spins $S$.
The transfer hopping integral ($t$) of the itinerant electron is
modified to  $t\cos({\frac{\theta}{2}})$ because of the double exchange
process which originates from strong Hund's
coupling between the spins of the core electrons and itinerant
electron \cite{gen}.
Here we would treat the core spins classically. For manganites
the core spins have $S=3/2$. However, for small values of J/t, 
the qualitative behavior of the phase diagram of the two-site 
Holstein-double exchange model does not depend on the
value of the spin or hopping as observed in Ref.\cite{Capon}.
Furthermore, D. M. Edwards and his group \cite{Edwards} pointed out 
that for such models the resistivity and transition temperature
$T_c$ do not vary much
with S, so that classical spin is a convenient approximation to
S=3/2 spins. Considering the out-of phase phonon mode which only couples
with the electronic degrees of freedom and treating the
spin classically, we obtain the MLF transformed Hamiltonian \cite{EPJB}
as
\begin{eqnarray}
\tilde{H_d}&=&\omega  d^{\dag} d +
(\epsilon_1- \epsilon_p) n_{1} + (\epsilon_2- \epsilon_p) n_{2}
-t cos(\frac{\theta}{2})
\nonumber \\
&& \left[ c_{1}^{\dag} c_{2}~ \rm{exp}(2 \lambda
(d^{\dag}-d))
+ c_{2}^{\dag} c_{1} \exp(-2 \lambda (d^{\dag}-d)) \right]
\nonumber\\
&+& \omega  (g_{+} -\lambda) (n_1-n_2) (d + d^{\dag}) +JS^2 cos\theta
\end{eqnarray}
In our previous work \cite{EPJB} we studied the above Hamiltonian
for a single polaron as a function of $e$-ph coupling for the
ordered case ($\epsilon_1$=$\epsilon_2$) using perturbation theory
with the variational MLF basis.  We found that the
nature of the ferromagnetic (FM) to antiferromagnetic (AFM)
transition as well as that of the polaronic state depends on
the relative values of $J$ and $t$. For small values of
$JS^2/t$  the magnetic transition does not coincide
with the polaronic crossover and a FM small polaronic state
exists between a large polaronic FM state and extremely
small polaronic AFM state. Similar phase diagrams have also been 
observed for both adiabatic and antiadiabatic limits in Ref. \cite{Capon}
for small values of $JS^2/t$. Here we will examine the effect of
site-energy disorder on such polaronic state and crossover.
We have followed the procedure of our previous work \cite{EPJB}
to find out the ground state properties of the double exchange
Holstein model with site disorder.

 In manganites the ratio of the site energy disorder potential
to the half band-width,
when calcium is doped in LaMnO$_3$, is about 0.4 \cite{Edwards}. 
The value of the disorder strength used in the present work is of that order.
It may be mentioned that previous studies \cite{Li} with a double 
exchange model have shown that the metal-insulator (MI) transition
due to off-diagonal disorder (associated with random spins in 
the paramagnetic phase) requires also a large diagonal disorder strength.
This led to the conclusion that disorder cannot alone account for
the MI transition in manganites and an $e$-ph coupling of intermediate 
range is needed for this purpose \cite{Edwards}.
 
In Fig. 7(a) and 7(b) we plot the angle ($\theta$) between the core spins,
the correlation function $\chi=\langle n_{1} (u_{1}-u_{2})
\rangle/2g_+\langle n_1\rangle$ and the kinetic energy ($E_{K.E}$)
as a function of $g_+$ to show the FM-AFM transition, polaron crossover
and the polaron-delocalization energy for both the ordered and disordered cases.
In presence of disorder, the FM-AFM magnetic transition and polaron
crossover occur at lower values of $g_+$. The reason behind this
behavior is that disorder effectively reduces the hopping, hence favors
formation of small polaron and the AFM phase.
In the AFM phase, the polarons are small polarons with almost vanishing
kinetic energy. However, the FM phase may have a large polaron character
or a small polaron character depending on values of $g_+$.
For low values of $t$ (=0.7), it is difficult to distinguish between the regions
of FM large polaron and FM small polaron since the change in $\chi$
is small and the curves for both $\chi$ and $E_{K.E}$ are very smooth
for disordered case. In Fig. 7(b) we consider
a higher value of $t$ (=1.5) with $JS^2/t=0.05$ as relevant for
manganite systems \cite{yu2}. Here a crossover from the FM large polaronic state
to a FM small polaronic state with reduced kinetic energy is
seen for the disordered case, before the transition to the AFM state.
So the site disorder does not make any drastic change to the qualitative
features of the ground-state properties of the double-exchange Holstein
model but smoothens the polaron crossover.
\begin{figure}[]
\centering
\includegraphics[scale = 0.5,angle=0]{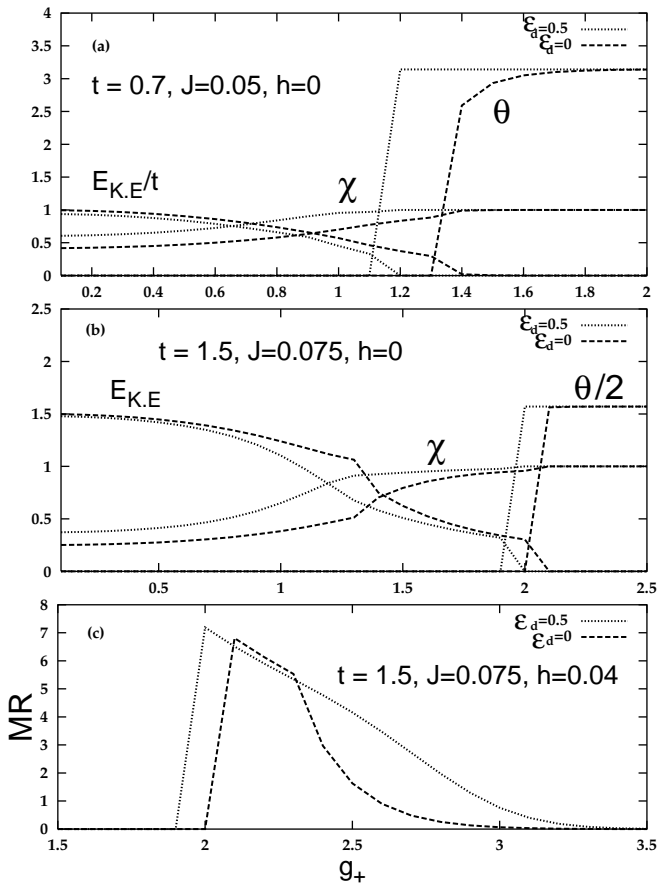}
\caption{Variation of $\theta$, $\chi=- <n_1(u_1-u_2)>
/2g_+ \langle n_1 \rangle$ and
$E_{KE}$ with $g_+$ for $\epsilon_d=0$ and 0.5 in absence of magnetic field
for (a) $t=0.7 $ and $JS^2=0.05$, and for (b) $t=1.5 $ and $JS^2=0.075$.
(c) The magnetoresistance MR=$(E_{KE}(h)- E_{KE}(0))/h$
as a function of $g_+$ for $t=1.5 $, $JS^2=0.075$  and $h$=0.04, $h$
represents the applied magnetic field.
}
\end{figure}
In Fig. 7(c) we plot the change in $E_{KE}$ due to the magnetic field ($h$)
as a function of $g_+$. This quantity may be related to the
magnetoresistance for a system in the thermodynamic limit as
pointed out in Ref. \cite {EPJB}.
In general, the $E_{KE}$ is a measure of delocalization of the
polarons and its change with the magnetic field gives field-induced
delocalization of the polaronic charge carriers. In Ref. \cite {EPJB}
we reported for the ordered case that for $JS^2/t$ =0.05 the change
in $E_{KE}$ due to the field has a broad peak around the
FM-AFM transition. We find that disorder makes the peak more broader
but the value of the change in $E_{KE}$ remains almost the same.
We believe that the magneto-resistance of similar model system in the
thermodynamic limit would show similar qualitative features.
We do not present here the results for larger values of $JS^2/t$
where the magnetic transition coincides with the polaron crossover
and the change in the $E_{KE}$ shows a sharp peak at the FM-AFM
transition \cite{EPJB} as we find that the site disorder does not change
that behavior. 

\vskip 1.0cm
\begin{center}
{\bf V. Conclusions}
\end{center}
\vskip 0.3cm

 To summarize, we have presented the results
on the two-site single polaron Holstein model in presence of
a site energy disorder $\epsilon_d$ which appears as a difference
in site energies. With increasing $\epsilon_d$ the retardation
between the electron
and associated deformation becomes weaker and the polaron crossover 
occurs at lower values of $e$-ph coupling. The polaronic kinetic 
energy is suppressed appreciably with disorder in the range from 
weak to intermediate couplings. But in the strong-coupling region
where the hopping through multi-phonon process is dominant, kinetic energy
is almost independent of $\epsilon_d$. For the oscillator wavefunction,  
a broad peak in strong-coupling
region is observed for the ordered case ($\epsilon_d$=0) in
addition to the main peak. In presence of disorder 
($\epsilon_d$ =1.0), this feature disappears.
We find that even a small disorder $(\epsilon_d$=0.001)
can localize the electron in presence of $e$-ph coupling. 
The polaron crossover precedes the localization in the 
non-adiabatic regime for low to intermediate values of disorder
($\epsilon_d \le t$).

 For the double exchange Holstein model both the magnetic transition 
and polaron crossover shift toward lower values of $e$-ph coupling
with increasing disorder strength. The qualitative features of 
the ground-state phase diagram does not change much with disorder.
The magneto-resistance for the model shows up in a broader region 
of parameter space in presence of disorder.

\end{document}